# Controlling the bi-photon orbital angular momentum eigen modes using asymmetric pump vortex beam

M. V. Jabir,* Ali Anwar, G.K. Samanta

*Photonic Sciences Lab., Physical Research Laboratory, Navarangpura, Ahmedabad, 380009, Gujarat, India.*
*Corresponding author: jabir@prl.res.in

**We report on controlling the bi-photon orbital angular momentum (OAM) eigen modes in the spontaneous parametric down conversion process by simply adjusting the asymmetry of the pump vortex beam. Adjusting the optic axis of the spiral phase plate (SPP) of phase winding corresponding to OAM mode, *l*, with respect to the beam propagation axis, we have transformed a Gaussian beam into an asymmetric vortex beam with OAM modes, *l*, *l*-1, *l*-2 ...0 with different weightages. Pumping the nonlinear crystal with such asymmetric vortices and controlling their asymmetry we have tailored the spiral spectrum of the bi-photon OAM eigen modes. Calculating the Schmidt number of the bi-photons we observe the increase in the spiral bandwidth of the OAM eigen modes and hence the dimensionality of the system. Although we have restricted our study to show the increase in spiral bandwidth of the bi-photons by simply controlling the asymmetry of the pump vortices, we can, in principle, further enhance the dimensionality of the entangled states by manipulating the pump beam size and crystal length.**

Quantum entanglement, the strong non-classical correlations in the joint measurement of two separate quantum systems, plays crucial role in variety of applications in quantum information processing, including quantum communication, quantum cryptography and teleportation [1-3]. So far, the quantum entanglement has been implemented mostly in photonic systems. The vast majority of the entangled photons are generated through spontaneous parametric down-conversion (SPDC) of high energy photon into low energy correlated photon pair [4]. With the first demonstration of entanglement with polarization [5], different degrees of freedom, such as spatial modes [6], time-energy [7] as well as continuous variables [8] have been used. Similarly, efforts have been made to produce photon pairs entangled between two different DoFs for example, polarization and spatial modes [9]. While entanglement in different DoFs have their own advantages and limitations, the entanglement in spatial modes e.g. orbital angular momentum (OAM) of photons has attracted great attention in recent times [6]. This is due to the fact that the OAM modes are discrete, integer valued, and form a theoretical infinite-dimensional Hilbert space essential for high dimensional entangled quantum states. The high dimensional entanglement not only increases the channel capacity [10], but also provides stronger violation of locality [11] and increases the security and robustness from eavesdropping.

The number of OAM modes generated in the SPDC process, commonly referred as the spiral bandwidth, dictates the OAM entanglement and the dimensionality of the system. The OAM modes of the SPDC process are commonly expressed in terms of bi-photon eigen modes of the paraxial OAM operators [12]. Mathematically the complete set of OAM eigen modes generated in SPDC process for the pump OAM mode of $l_p$ are represented as, $\sum_{l_p=-\infty}^{\infty} \sum_{l_i=-\infty}^{\infty} C_{l_i} |l_i\rangle |l_p - l_i\rangle$. Here, $l_i$ and ($l_p - l_i$) are the idler and signal OAM modes, respectively, generated in SPDC process. While efforts have been made to understand the effect of pump profile and phase matching on bi-photon eigen modes [12, 13], the majority of such studies have dealt with the OAM entanglement using with Gaussian pump, $l_p$=0. In such cases, the bi-photon eigen modes represented as, $\sum_{l_i=-\infty}^{\infty} C_{l_i} |l_i\rangle |-l_i\rangle$ span a single subspace in the OAM Hilbert space. While manipulation of pump beam waist radius and the crystal parameters can increase the spiral bandwidth of the OAM mode, however, the bi-photon eigen modes are restricted to the same subspace. Similarly, one can manipulate the spatial mode of the pump beam and directly generate the spatially entangled states [9, 14]. However, complete access to the bi-photon eigen modes, $\sum_{l_p=-\infty}^{\infty} \sum_{l_i=-\infty}^{\infty} C_{l_i} |l_i\rangle |l_p - l_i\rangle$ in the OAM Hilbert space require pump beam to carry superposition of OAM modes.

Optical vortices, having the phase distribution represented as exp($il\theta$), where $\theta$ is the azimuthal angle and the integer $l$, known as vortex order or topological charge of the vortex, carry OAM of $l\hbar$ per photon. Typically, the optical vortices are produced through the spatial mode conversion of Gaussian laser beams using mode converters including spiral phase plates (SPPs), q-plates and the computer generated holography technique based spatial light modulators (SLMs). Most of these mode converters are designed to generate a single valued OAM mode, $l$ [15]. On the other hand, fractional optical vortex beam contains large number of integer OAM modes with different weightage. Recently, it is observed that the off-axis vortex (asymmetric vortex) beams have broad OAM modal distribution [16, 17]. Using such asymmetric optical vortex beam, here, we investigate the OAM eigen modes of the bipartite system generated in SPDC process. Using SPP and simply shifting its optic axis with respect to the beam propagation axis, we have generated pump beam with a broad spectrum of OAM modes of different weightages. Pumping the nonlinear crystal with such asymmetric vortices we directly transferred the pump OAM spectrum to the down converted photons producing bi-photon eigen modes spanned over more number of subspace in the OAM Hilbert space. Calculation on Schmidt number [18] shows that the increase in the OAM modes with increase in asymmetry in the pump vortices. Additionally, shifting the SPP we can tune and access any particular subspace of the OAM Hilbert space. To the best of our knowledge, this is the first experimental report on controlling the OAM eigen modes of the bi-photon state spanned by the pump OAM modes.

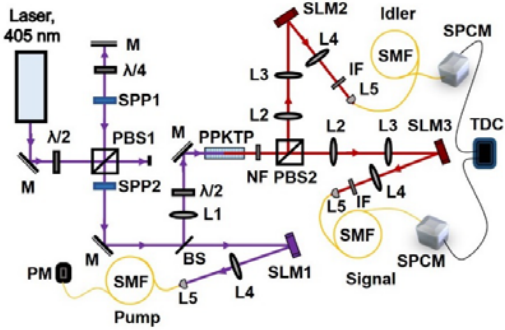

Fig. 1. Schematic of the experimental setup. $\lambda/2$, $\lambda/4$, half-wave plate and quarter wave plate at 405nm; PBS, polarizing beam splitter cube at 405 nm and 810 nm; SPP1-2, spiral phase plate; M, mirrors; L1-5, lenses; PPKTP, nonlinear crystal in a temperature oven; SLM, Spatial light modulator; IF, interference filter; NF, notch filter; SMF, single mode fibre; SPCM, single photon counting module; TDC, coincidence counter; PM, power meter

The schematic of the experimental setup is shown in Fig. 1. A continuous-wave, single-frequency (line-width <12 MHz), diode laser of 100 mW output power at 405 nm in Gaussian spatial profile ($TEM_{00}$ mode), is used as the pump source. The combination of a half-wave plate ($\lambda/2$) and polarising beam splitter (PBS1) cube controls the laser power to the experiment. Using the spiral phase plates, SPP1 and SPP2, having phase-winding numbers, $m=1$ and 2, respectively, and the vortex doubler setup [15], comprised of PBS1, quarter-wave plate ($\lambda/4$) and a mirror (M), we have converted the Gaussian pump beam into vortex beam of orders (OAM modes), $l_p=1$ to 6. The vortex beams have symmetric intensity distribution and carry OAM mode, $l_p$, equal to the phase winding numbers, $m$, of the SPPs when the optic axis of the SPP coincide with the beam propagation axis. However, any misalignment in terms of tilt and or displacement between the beam propagation axis and the optics axis of the SPPs create asymmetry in the intensity distribution of the generated vortex beam. The resultant beam carries a broad OAM mode spectra and the distribution of the mode spectra vary with the asymmetry of the vortex beam. A fraction of the pump beam intensity transmitted through the beam splitter (BS) is used to study pump OAM modes.

Using the lens, L1, of focal length, $f_1=750$ mm, the pump beam is focused at the centre of a 30 mm long, 2×1 mm$^2$ in aperture, periodically poled potassium titanyl phosphate (PPKTP) having single grating of period, $\Lambda=10$ μm, for type-II ($e \to o+e$) phase-matched degenerate, collinear down conversion of pump beam in to 810 nm. Since the degenerate photons have orthogonal polarization states, we have considered the photons in horizontal and vertical polarization states as signal and idler, respectively. The notch filter (NF) removes the undepleted pump from the signal and idler. We have separated the signal and idler in transmitted and reflected ports, respectively, of the PBS2. Using the lenses, L2 and L3 of focal lengths, $f_2=100$ mm and $f_3=500$ mm in $2f_2$-$2f_3$ combination we imaged the signal and idler photons with 5X magnification on the surface of SLM2 and SLM3, respectively. The OAM content of the pump, signal and idler are measured using the projective technique [20], where, the incident photons are diffracted by the SLM (SLM1, SLM2 and SLM3) having blazed fork grating of different OAM orders. The first order diffracted photons are coupled to a single mode fibre by re-imaging the SLM plane with the lens combination, L4 and L5 ($f_4=750$ mm and $f_5=4.6$ mm), in $2f_4$-$2f_5$ configuration and subsequently measured using a detector. A power meter (PM) measures the power distribution of the pump OAM modes. The single photon counting modules (SPCMs) and the time to digital convertor (TDC) measure the coincidence of the signal and idler photons. We have used a coincidence window of 8.1 ns. The interference filter (IF) has a transmission bandwidth of 2.1 nm centred at 810nm.

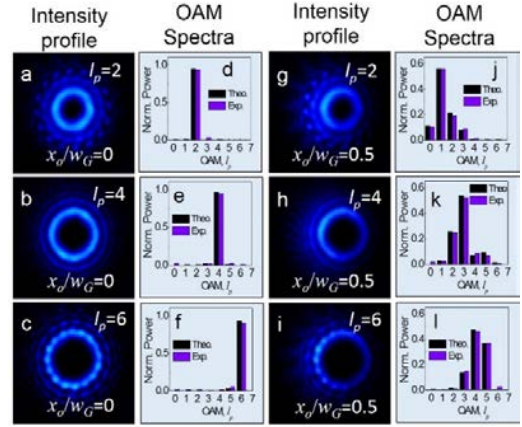

Fig. 2. (a-c) Far-field intensity distribution, corresponding (d-f) OAM spectra of symmetric ($x_o/w_G=0$), and (g-i) intensity distribution, corresponding (j-l) OAM spectra of the asymmetric ($x_o/w_G=0.5$) pump vortex beams of orders, $l_p=2$, 4, and 6. The black and violet bars represent the theoretical and experimental results respectively.

We have studied the OAM spectra of the asymmetric vortex beam by measuring the far-field intensity profile and corresponding OAM modes of the pump vortex beam with the results shown in Fig. 2. To create the asymmetric position of the dark core of the vortex beams, we have moved the SPPs with respect to the beam axis with a normalized distance, $x_o/w_G$ [17]. Here, $x_o$ is the shift of the optic axis of SPP away from the beam propagation axis and $w_G=1.2$ mm, is the full-width at half-maximum (FWHM) radius of the Gaussian beam at the SPP plane. Using SPP1 and SPP2 in the vortex doubler setup and adjusting the optic axes of the SPPs with respect to the beam propagation axis so that $x_o/w_G=0$, we have recorded the intensity profile of the vortex beams. As evident from first column, (a-c), of Fig. 2, the vortex beams have symmetric intensity distribution with increasing dark core size. We have measured the OAM modes of the symmetric vortex beams, as shown in the second column, (d-f), of Fig. 2, to be $l_p$=2, 4, to 6, respectively. However, as evident from the third column, (g-i), of Fig. 2, the shift in the SPPs position by a normalized distance of, $x_o/w_G$=0.5, results asymmetric intensity distribution of the vortex beams with the dark core moved away from the beam propagation axis. Similar to our previous report [17], the increase in shift distance, $x_o/w_G$, pushes the dark core of the vortex beam away from the centre of the beam resulting larger asymmetry in the beam. Since the azimuthal phase variation in the beam is no longer uniform with respect to the beam axis, one can expect the beam to carry a distribution of OAM modes. Fourth column, (j-l), of Fig. 2, shows the measured OAM mode distribution of the asymmetric vortex beams generated for, $x_o/w_G$=0.5. It is evident from Fig. 2, that the shift of SPPs from $x_o/w_G$=0 to 0.5, transform the OAM modes, $l_p$=2, 4, and 6 to a distribution of OAM modes of (0, 1, 2, and 3), (2, 3, and 4) and (3, 4 and 5), respectively. Using the mathematical formula of Ref. [20] we have calculated the OAM mode distribution (black bars) of the asymmetric vortex beams for our experimental conditions and found a close agreement with our experimental results (violet bars). As predicted [19] and experimentally verified [17], here, we also observe the variation in OAM mode distribution of the asymmetric vortex with $x_o/w_G$. For very high asymmetry, $x_o/w_G$>>1, one can reach to the Gaussian mode. Using this technique, it is evident that one can control the OAM mode distribution of a vortex beam of OAM mode, $l_p$, into the mixture of OAM modes, $l_p$, $l_p$-1, $l_p$-2,…,0 of different weightages by simply shifting the optic axis of SPPs away from the beam propagation axis.

Fig. 3. (a-c) Distribution of OAM modes of asymmetric pump beam of $x_o/w_p$= 0, 0.75 and 1.25, corresponding spiral spectrum for bi-photon OAM eigen modes (d-f), measured and (g-i) simulated. (j-l) A portion of

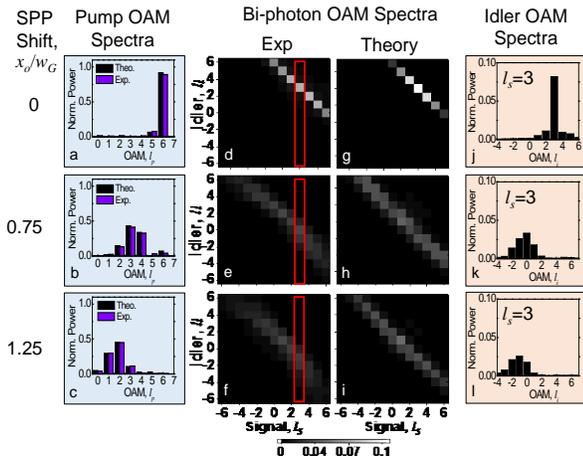

the spiral spectrum of idler at a fixed OAM mode ($l_s$=3) of the signal.

Knowing the OAM mode distribution of the asymmetric vortices, we have studied the transfer of OAM modes of pump photon to the signal and idler photons in parametric down conversion process. Pumping the PPKTP crystal at temperature T=47.5°C with asymmetric pump vortex beams at 405 nm, we have measured the OAM modes of signal and idler photons by using projective measurement technique. To verify the reliability of the projective measurement setup, we first have measured the OAM bi-photon eigen modes by using Gaussian pump beam. Using a CCD camera we have estimated the beam waist radius of the Gaussian pump beam at crystal to be, $w_p$ ~60 µm. Similarly, using back alignment technique [20], we have measured the waist radius of the signal and idler beams to be $w_s$, $w_i$ ~60 µm ensuring the ratio, $\gamma = w_p/w_i$ ~1, at the crystal plane to ensure optimum collection of the generated photons. Projecting the signal and idler photons in different OAM modes and measuring their coincidence counts we have measured the bi-photon OAM eigen modes for the Gaussian pump beam and consequently estimated the Schmidt number. Using the mathematical expression of the Schmidt number, $K = \beta\left(\frac{w_p^2 + 4\alpha^2 b^2}{4w_p \alpha b}\right)^2$, of Ref. [21], where, $\alpha = 0.85$, $\beta = 1.65$ $w_p$, and $k_p$ are the beam waist radius and the wave vector of the pump respectively, and $L$ is the length of the crystal, we have calculated the Schmidt number to be, $K_{theo}$ ~2.82, close to the measured value of, $K_{exp}$~2.86. Such close agreement between the experimentally measured Schmidt number to that of the theoretical Schmidt number confirms the reliable measurement capability of our OAM detection system.

Fig. 4. Variation in the measured value of Schmidt number, $K$, as a function of asymmetry in the pump vortex beam.

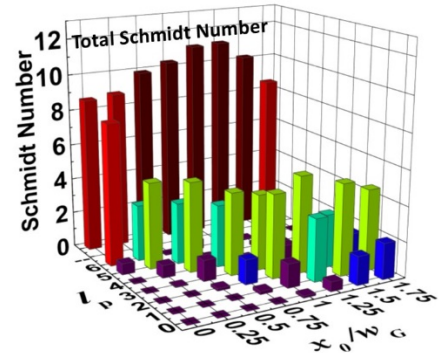

Using the same detection system, we pumped the crystal with vortex beam of order, $l_p$=6, with varying $x_o/w_G$ value and measured the bi-photon OAM eigen modes of the down converted photons. The results are shown in Fig. 3. As evident from the first column, (a-c), of Fig. 3, the pump OAM spectra is varying from a single mode, $l_p$=6 at $x_o/w_G$ =0 to a distribution of OAM modes, $l_p$=4, 3 and 2 at $x_o/w_G$ =0.75, and $l_p$=3, 2 and 1 at $x_o/w_G$ =1.25 with different weightages. From the second column, (d-f), of Fig. 3, it is evident that the bi-photon OAM eigen modes belong to the subspace spanned by OAM mode of, $l$=6, for the pump beam of single OAM mode, $l_p$=6, satisfying OAM conservation in down conversion process. However, with the increase in asymmetry of the pump vortex beam resulting from $x_o/w_G$=0, 0.75 and 1.25 we observe the increase in the span of bi-photon OAM eigen modes. Such broadening in the OAM eigen modes of the bi-photons can be understood as follows. As the OAM modal distribution of the pump beam broadens with the beam asymmetry, the generation of bi-photon OAM eigen mode by the constituting pump OAM modes of the asymmetric vortex beam result in the broadening of spiral spectrum. Using our experimental parameters, we have simulated [22] the bi-photon OAM eigen modes with results shown in third column, (g-i) of Fig. 3, in close agreement with our experimental results. To gain further insight on broadening of spiral bandwidth of the down converted photons, we have analysed the OAM mode distribution of the idler photons for a fixed signal OAM mode of $l_s$=3 [see the red boxes of Fig. 3 (d-f)]. The results are shown in fourth column, (j-l), of Fig. 3. We can clearly see the broadening of the OAM eigen modes of the bi-photons, a close resemblance to the OAM spectrum of the asymmetric pump vortices. It is also evident from the present study that one can tune the spiral bandwidth of the bi-photon eigen modes by simply adjusting the optic axis of the SPPs with respect to the beam propagation axis.

As the pump OAM spectrum vary with its asymmetry and the pump OAM mode, $l_p$, spans the OAM eigen-modes of the bi-photons, we have studied the effect of pump beam asymmetry on the Schmidt number, $K$. Using the pump vortex order, $l_p$=6, we have calculated the Schmidt number for OAM modes, $l_p$=1 to 6 with asymmetry parameter $x_o/w_G$. The results are shown in Fig. 4. As evident from the Fig. 4, the pump vortex beam with asymmetry, $x_o/w_G$ =0 has OAM mode of, $l_p$=6, and a total calculated Schmidt number of, $K$=8.05 with major contribution from the pump OAM mode, $l_p$=6. However, with the increase of asymmetry, $x_o/w_G$, from 0 to 1.75, we observe a varying distribution of pump OAM modes, $l_p$, $l_p$-1, $l_p$-2,..,0 of different weights, and the corresponding Schmidt numbers. It is also interesting to note that the total Schmidt number, the sum of the Schmidt numbers of individual OAM modes, $l_p$, varies from 8.68 to 8.52 for the increase of $x_o/w_G$ from 0 to 1.75 with a maximum of 11.01 at the pump beam asymmetry of, $x_o/w_G$=1. Such increase in the total Schmidt number, a measure of total number of bi-photon OAM eigen modes, can be attributed to the broadening of the OAM spectra of the pump beam near the asymmetry of, $x_o/w_G$=1.

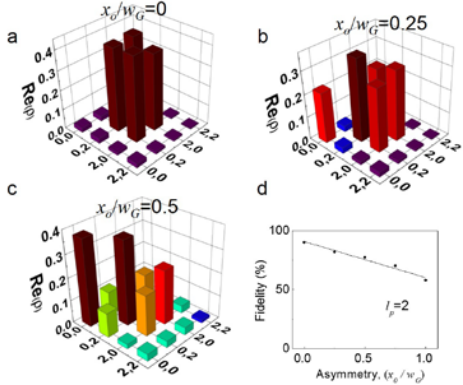

Fig. 5. (a-c) Variation of the density matrix of two dimensional OAM entangled state with the asymmetry, $x_o/w_G$=0, 0.25 and 0.5 of pump vortex OAM mode, $l_p$=2. (d), The change in the fidelity of the state with pump beam asymmetry.

Further, we have studied the effect of vortex asymmetry on the quality of two dimensional OAM entangled states by performing the state tomography for different values of asymmetry parameter, $x_o/w_G$. The results are shown in Fig. 5. Pumping the crystal with vortex beam of OAM mode, $l_p$=2, we have generated the Bell state, $|\psi\rangle = \frac{1}{\sqrt{2}}(|2,0\rangle+|0,2\rangle)$. The Fig. 5 (a)-(c) show the real parts of the density matrix of measured Bell state for the pump vortex asymmetry, $x_o/w_G$=0, 0.25 and 0.5, respectively. It is evident from Fig. 5(a)-(c) that the density matrix of the Bell state change with the increase in asymmetry of the pump vortices from $x_o/w_G$=0, 0.25 and 0.5. To get further insight, we have measured the Bell state fidelity for different asymmetries in the pump vortices with the results shown in Fig. 5(d). It is evident from the Fig. 5(d) that the fidelity of the Bell state decreases with the increase in asymmetry of the pump vortex beam. Such decrease in the fidelity of the Bell state can be attributed to the cross talk among different modes of the higher dimensional OAM entangled state. One can, in principle, obtain high fidelity by considering higher dimensional bases in the measurement.

In conclusion, we have tailored the OAM mode distribution of a vortex beam from the single mode to a mixture of OAM modes by controlling the asymmetry in its intensity distribution. Using such tailored pump OAM spectra we have studied the bi-photon OAM eigen modes in the down conversion process. The direct correlation of the pump OAM spectra to the bi-photon OAM eigen modes in SPDC process confirms the direct control of spiral bandwidth of the entangled photons. We have quantified the broadening of the bi-photon OAM eigen modes by calculating the Schmidt number, and observe the enhancement in the dimensionality of the bi-photon OAM eigen modes by tailoring the asymmetry of the pump vortices. This generic method may be utilized for increasing the information capacity of various quantum information protocols.